\documentstyle[aps,twocolumn,psfig]{revtex}   

\begin{document}
\title{Nonequilibrium transitions induced by multiplicative noise}
\author{Walter Genovese$^{1}$, Miguel A. Mu{\~{n}}oz $^{1}$,
 and J.M. Sancho $^{2,3}$}
\address{ $^{1}$ Dipartimento di Fisica,
Universit\'a di Roma ``La Sapienza'',~P.le
A. Moro 2, I-00185 Roma, Italy}
\address{$^{2}$ Institute for Nonlinear Science, Dept.0407,
University of California, San Diego
9500 Gilman Drive, La Jolla, Ca 92093-0407. USA
}
\address{$^{3}$
and Depatament d'Estructura i Constituents de la Mat\'eria,
Av. Diagonal 647, 08028 Barcelona. Spain.
 }
\date{\today }
\maketitle

\begin{abstract}
  
   A new simple model exhibiting a
 noise-induced ordering transition (NIOT) and
 a noise-induced disordering transition (NIDT),
 in which the noise is purely 
multiplicative, is presented. Both transitions are found in two 
as well as in one dimension (where they had not been previously 
reported).
We show analytically and numerically
 that the critical behavior of these two
transitions is
described by the so called
multiplicative noise(MN) universality class. A computation of the 
set of critical exponents 
 is presented in both $d=1$, and $d=2$ (where they have not been
previously measured).

\vspace{8pt}
\vspace{8pt}

\end{abstract}

\begin{text}
\narrowtext

\section{Introduction}

It is well established that the effects of noise in
stochastic
equations can be rather surprising and counterintuitive. The presence
of a noisy term can modify in a quite an unexpected way the behavior
of the corresponding deterministic (noiseless) equation.
 A well known example of this is the {\it stochastic
resonance} in which the response to an external periodically oscillating
 field 
of a stochastic 
system 
 is enhanced
by increasing its internal noise  amplitude \cite{sr}.
 Other examples
are the resonant activation \cite{sa}, 
noise-induced spatial patterns \cite{n1}, the noise induced 
ordering transition (NIOT) 
\cite{Raul,BK,joan}, and
the noise induced disordering transition (NIDT) 
 \cite{Raul}. This two last transitions were
 first analyzed together in an interesting work by Van den
Broeck, Parrondo and
Toral \cite{Raul} (VPT hereafter).
 They considered a Langevin equation with a monostable
 deterministic term, 
to which a non-trivial noise is added. Owing to 
the combined effect of the noise and the spatial coupling two
real phase transitions were shown to appear:
 for small noise amplitudes the system is  
disordered, while
  by increasing the noise
amplitude the system  gets ordered in a symmetry-breaking state
(exhibiting therefore a NIOT). 
If the noise is increased further the ordered state is destroyed, 
 (and therefore there is a NIDT).
    Afterwards, different works on NIOT and NIDT
 appearing
in different models and systems have been proposed in the literature
\cite{Sancho,Kim,Muller}. In particular, in \cite{Sancho}, a 
noise-induced reentrant transition
is reported for a time dependent Ginzburg-Landau
model with both additive and multiplicative noise.
 In \cite{Kim,Muller} first order transitions are shown to be
induced by the presence of non-trivial noises.

  An interesting task is that of understanding the nature of this
striking phenomena in the simplest possible model: understanding
whether it comes from an interplay between additive and multiplicative
noises or not, and determining
whether the critical behavior of these transitions 
 can be described by
any standard universality class.
  In this direction, the following Langevin equation 
has been proposed by VPT as a possible simplest model exhibiting
these two noise induced transitions:

\begin{equation}
{\partial \phi (x,t) \over \partial t} =  F[\phi(x,t)] + D\nabla^2 \phi +
 G[\phi(x,t)] \eta(x,t)
\label{vpt}
\end{equation}
with
\begin{equation}
 F[\phi(x,t)]= - \phi (1 + \phi^2)^2,   ~~~~~
 G[\phi(x,t)]=  (1+ \phi^2) 
\label{fg}
\end{equation}
where $\phi$ is a field, 
$\eta$ is a white Gaussian noise with $<\eta(x,t)>=0$ and
$ < \eta(x,t) \eta(x',t')> = \sigma^2 \delta(x-x') \delta(t-t') $,
and the
Stratonovich interpretation is considered \cite{Gardiner}.
  Although preliminary simulations seemed to indicate
 that both the
NIOT and the NIDT exhibit
exponents compatible with
mean-field like values \cite{Raul}, it has been recently elucidated that
both transitions belong into the kinetic Ising model
 universality class 
\cite{Geoff}.
  In particular, very extensive simulations and finite size analysis 
have been performed
 \cite{Kawai} showing that all the measured exponents 
are compatible with  their corresponding Ising values.
 The same conclusion can also be reached by
field theoretical analysis \cite{Geoff,Ind}; and it is also extensible
to the model of Ref.\cite{Sancho}  with
both additive
and multiplicative independent noises.
  In particular, a naive
power counting argument permits to conclude that all the
terms in eq. (\ref{fg}) others than the linear, the cubic, and
the Laplacian term 
in the deterministic part, and the constant term in the noise,
are {\it irrelevant} in the renormalization group sense. In this
way, the relevant part of eq. (\ref{fg}) coincides with
the Hohenberg and Halperin model A \cite{HH}, and consequently
 eq. (\ref{fg}) is in the kinetic Ising model universality class.
 The irrelevant terms, i.e.,
 the $\phi^5$ term in $F(\phi)$ and the 
$\phi^2$ term in $G(\phi)$,
 play a key role in determining the structure of
the phase diagram, but do not affect the critical behavior. 

   In this paper, we face the two previously arisen issues,
namely,
we propose a new candidate with only multiplicative noise 
exhibiting these two noise transitions,
and study their critical behavior under a new
perspective. 
   We will conclude that the phenomena is purely
 multiplicative, and therefore the presence of additive noise
(i.e. the constant term in $G(\phi)$) is not required 
to generate neither the NIOT nor the NIDT. 
As a byproduct we show that the conclusion reached in the 
previous works \cite{Kawai} that both transitions are
in the kinetic Ising model universality class is not general.  
We will show that in the {\it minimal model} we propose, that 
reproduces the
phenomena under discussion, the transitions belong into a new
recently proposed universality class: the so called 
{\it multiplicative noise (MN) universality class} 
\cite{Noi1,Pik}.

  \section{The model}

  \subsection{Motivation}

  A physically intuitive justification of the NIOT
 has been recently proposed \cite{Kawai}.
It is based on the following observation: the short time evolution
of the average value of the field, $\phi$, of a generic 
Langevin equation of the form (\ref{vpt}) with generic
forms of the functionals $F$ and $G$,  is easily
found to be given by:
\begin{equation}
{\partial <\phi> \over \partial t} = F(<\phi>) +  {\displaystyle
 \sigma^2 \over 2} G(<\phi>) G'(<\phi>)
\end{equation}
 when
 $D=0$.
  
 It is shown in \cite{n1,BK,Sancho,Kawai} that for the models
exhibiting 
a NIOT  the previous 
equation has a linear
 instability for the homogeneous state $<\phi>=0$;
this means that the average
value of  $\phi$ grows at initial times, while it would
decrease monotonously towards $0$ in absence of the
noise term $G(<\phi>) G'(<\phi>)$, i.e., in absence of
the multiplicative part of the noise, which is the only one
giving a non-vanishing contribution
to $G'(\phi)$. 
  When the spatial coupling is turned on , (i.e., $D \neq 0)$),
 it favors
neighbor sites to take similar values of the field, and, consequently,
the $<\phi> \neq 0$ solution, induced by the
multiplicative noise, is stabilized. This is the physical mechanism 
at the origin of the noise-induced phase transition.
At this point we want to stress the fact that is the multiplicative
part of the noise the one responsible of the generation
of the described short-time instability, and therefore of the NIOT.

On the other hand, the NIDT has a more conventional interpretation,
analogous to standard equilibrium transitions:
noise (temperature) destroys long range order, thus restoring the symmetry
of the homogeneous state.

   Inspired by the previous considerations we propose 
a simpler model to analyze the noise-induced transitions
in which the noise is {\it purely multiplicative}. In all the 
previously studied models an additive noise term was also present:
 it is not essential for the phenomena but it does change  
the universality class of the two transitions.

    \subsection{Definition}

   Our model is defined by a Langevin equation (\ref{vpt}) with
the functionals $F$ and $G$ given by:
  \begin{equation}
 F[\phi]= - a \phi - \phi^3,  ~~~~
 G[\phi]=  \phi (1-\alpha \phi^2)^{1/2}
\label{our}
\end{equation} 
 where the Gaussian noise, $\eta$, is  specified
by $<\eta(x,t)>=0$ and 
 \begin{equation}
 < \eta(x,t) \eta(x',t')> = \sigma^2 
 \delta(x-x') \delta(t-t'),
\label{our2}
 \end{equation}
and $\phi$ is a positive-definite field, and the Stratonovich interpretation
is considered.

   In this way, we have a very simple deterministic term and
 a purely multiplicative noise, i.e., with no additive part.
  By setting the parameter $\alpha$ to zero, we get the NIOT,
 therefore the more interesting 
phenomena under study is present in the simple 
multiplicative noise Langevin equation studied in \cite{Noi1},
(that is nothing but our model with $\alpha=0$).
Nevertheless, that model does not exhibit a NIDT
due to the fact that the order parameter keeps on increasing monotonously
with increasing noise amplitude. In order to recover the NIDT
a number of possible mechanisms can be invoked. Among them we have
chosen to introduce the term in eq. (\ref{our2}) parametrized by $\alpha$. 
This ''trick''  keeps the order parameter
from growing indefinitely for large values of $\sigma$. Let us point
out that eq. (\ref{our}) cannot overpass the upper limit $\phi=\alpha^{-1/2}$.
 Observe that while the multiplicative noise is essential to
generate the NIOT, the NIDT is reproduced in a somehow artificial;
the NIDT is
     more naturally induced by standard
additive noise, but it also changes the critical behavior.

     \subsection{Renormalization Group results}

  Even though we do not present here the details  of the calculations,
it is a straightforward task to conclude from a naive power counting
analysis that this model belongs into the so called multiplicative
noise (MN) universality class  (defined by our model with $\alpha=0$,
see \cite{Noi1}).
  For that, it has only to be noticed 
that $\alpha$ is an irrelevant parameter in the renormalization group sense.
 This conclusion is not
altered when introducing diagrammatics corrections. Therefore
the critical exponents in all the possible transitions exhibited
by this model are expected to be in the MN universality class \cite{Noi1}.

     \subsection{Mean field results}

    We have first performed a mean field analysis of the model.
For that we consider a given point $x$ and assume its nearest neighbors
to take a constant undetermined value, $m$, the 
 expectation value of $\phi$ is calculated as a function, $f$ of $m$,
$<\phi>=f(m)$, and the equation is closed self-consistently, i.e,
$<\phi>=m$. 
 In figure 1 we observe that, for small values of
$\sigma$, $y=f(m)$ does not intersect
 with the straight line $y=m$, while they do 
intersect each other
for values of $\sigma$ larger than a certain critical value, $\sigma_{c1}$.
The intersection
point is very small
 nearby $\sigma_{c1}$, grows while increasing the noise amplitude up to a 
point in which starts decaying
 towards zero for large $\sigma$ values (in a extremely 
slow fashion; see figure 1). We have shown analytically
 that in the infinite noise limit the only solution
is $m=0$ for $\alpha \neq 0$.
 Therefore,
in this approximation the model displays 
a NIOT and a NIDT, and both of them are
continuous.  

  The results for the $\alpha=0$ are also displayed in figure 1(a). Observe 
that as it was said previously the order
 parameter increases monotonously with $\sigma$
and therefore the model exhibits a NIOT but not the NIDT in mean field
approximation.

\begin{figure}
\centerline{\psfig{figure=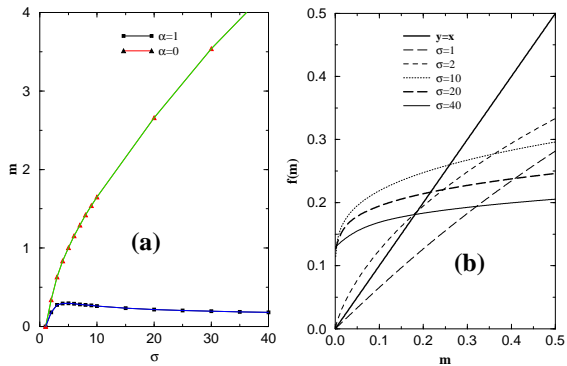,width=7cm}}
\caption{ (a) Mean field value of $<\phi>$ for
different values of $\sigma$, for $\alpha=0$ and 
$\alpha=1$. 
(b)  
 Curves $f(m)$ plotted
 versus $m$ for different values of $\sigma$ and $\alpha=1$. 
The mean field solution plotted in (a) is given by the intersection point
of the different curves  
with the straight line, $y=m$ (see text).   
}
\end{figure}

      \subsection{Numerical results}

   We have studied the previous model in both one and two dimensions
 by integrating numerically the associated Langevin equation.
For that we have used a Runge-Kutta
type of method (known as the Heun
method \cite{NR,Max}). The time step is chosen to be $\delta t=
10^{-3}$, and other parameters fixed in the simulations are :
 $a=1$, $D=0.2$ for the one-dimensional simulations,
and $D=1$ in two dimensions.
 Periodic boundary conditions are
employed, and lattice sizes up to $L=1000$ in $d=1$, and
$ L=40 $ in $d=2$.
    The main results we find are as follows:

   1-  We find numerically the two  noise-induced transitions in
both
$d=1$ and $d=2$ for $\alpha=1$ (for $\alpha=0$ we find only
the NIOT; in the ordered phase the order parameter grows
monotonously  with increasing $\sigma$ and therefore there 
is no NIDT).
 In figure 2 the order 
parameter, $<\phi>$ is plotted as a function of the noise amplitude,
$\sigma$, for  $d=1$ and $d=2$ (with $\alpha=1$).
Observe that due to the high values of $\sigma$ the finite size effects
are much more pronounced
in the NIDT than in the NIOT, and therefore
much larger systems and better statistics are required to determine
accurately the critical point and the associated critical behavior of the
NIDT.
     
 Let us stress the fact that we find a pair of phase transitions
even in $d=1$. This is not expected to be the case in any of
the previously studied models due to the fact that 
in all of them the constant additive part of the noise "cohabits" with
the purely multiplicative term, and that changes dramatically 
the critical behavior, rendering the system Ising-like, and as 
is well established,
systems of that nature cannot get ordered in $d=1$.

\begin{figure}
\centerline{\psfig{figure=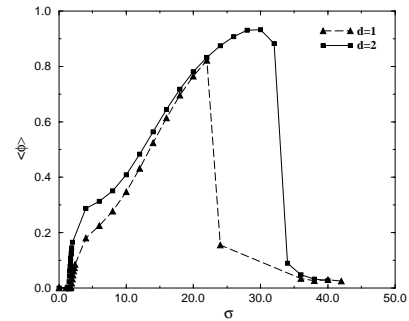,width=6cm}}
\caption{ Stationary expectation value of $\phi$ as a function
of the noise amplitude $\sigma$ in one and two dimensions. 
}
\end{figure}

  2- We have performed a detailed numerical analysis  
 of the NIOT in one and two dimensions. 
 We have computed the critical exponents defined by:
\begin{eqnarray}
  <\phi (t \rightarrow \infty) > & \sim & | \sigma -\sigma_{c1}|^{\mu} 
 \label{mu} \\
  <\phi (t, \sigma=\sigma_{c1}) >  & \sim  & t^{-\theta}
\label{theta} \\
  <\phi^n (t \rightarrow \infty, \sigma=\sigma_{c1}) >
 & \sim & | a - a_{c1} |^{\beta_n} 
  \label{beta}. 
\end{eqnarray}
    
In order to locate accurately the critical point and determine the 
exponent $\mu$ in $d=1$ we
 looked for the value of $\sigma_{c1}$  for which the best linear
fit is got in a plot of
 $ log(<\phi>)$ 
versus $log(\sigma-\sigma_{c1})$. The exponent $\mu$ is given by
 the slope of such a plot with
the optimized value of $\sigma_{c1}$.
  We find:
 $\sigma_{c1}= 1.80 \pm 0.02$ with $\mu=0.8 \pm 0.1$ in $d=1$.

Afterwards, we determine $\theta$
from the asymptotic decay of $<\phi(t)>$ right at the critical
point, and get $\theta=1.0 \pm 0.1$ in $d=1$.
     Next we fix the noise amplitude to its critical value and
perform simulations changing the value of the linear term $a$. 
By doing that we find: $\beta_1=1.50 \pm 0.05$ and
 $\beta_2=1.7 \pm 0.2 $.
  
  Let us point out that the exponents $\theta$, $\beta_1$ and $\beta_2$
 are in very good agreement with
their corresponding values in the one-dimensional 
multiplicative noise universality
class \cite{Noi1,DNA},
 and the predicted scaling relations hold.
 In particular,
 we are tempted to conclude from our numerical simulations
 that $\beta_1=\beta_2 =3/2$  and $\theta=1$ in $d=1$
 and, in this way, all the
exponents in the MN universality class would take rational values. 
 On the other hand, $\mu$ is a new exponent that had not been 
previously determined for the MN universality class \cite{note}.

   In $d=2$ concluding results are much tougher to obtain.

 We have verified that the 
behavior of the computed magnitudes at the critical point is compatible
with exponents $\mu=1.03 \pm 0.05$, $\theta=1.75 \pm 0.30$,
 $\beta=1.14 \pm 0.05$, and $\beta_2=1.70 \pm 0.10$ \cite{Noi1,future}).
 A more detailed discussion
of this point will be presented elsewhere \cite{future}.
 Up to our knowledge this is the first computation
of the critical exponents of MN in $d=2$.

  3- We have also performed a study of the  
NIDT in $d=1$ and $d=2$.
 As was said previously, finite-size effects 
are much more severe in this transition than they are
 in the NIOT.
 For that reason, a  
finite-size-scaling analysis has to be performed in order to elucidate
its universality class. This analysis will be 
published elsewhere \cite{future}. In any case,
based on both theoretical 
 and preliminary numerical analysis
we believe that these transitions also belong into the MN universality
class \cite{first}.
 In particular, the fact that the NIDT is present
in $d=1$ excludes the possibility of Ising-like behavior, and
relevance arguments strongly support the previous hypothesis.

\section{Conclusions}

We have introduced  a very simple stochastic model 
exhibiting two noise-induced transitions. 
We  have
shown
that the noise-induced ordering 
transition (NIOT) is essentially originated by a purely
 multiplicative noise,
and have found it  in $d=1$ (where it has not been previously 
reported) and 
 $d=2$ (where we measure the MN exponents for the first time).
 That shows
that the multiplicative noise is capable to originate either a NIOT 
or a NIDT even in that low dimension in which
 phase transitions are rare.

 We have verified both analytically and numerically 
that NIOT and NIDT of our model belong into the 
recently elucidated multiplicative-noise (MN) universality class
\cite{Noi1}. 
 The result, obtained in previous works, that
the NIOT and the NIDT 
 belong into
the kinetic Ising model universality class is not general;
that result derives from the fact that an additive noise term is
also present in those models, and that type of term, unnecessary
as we have shown to generate the phenomena under study, does change
the system critical behavior, rendering the universality class
model dependent.

\section{Acknowledgements}  
    It is a pleasure to acknowledge J. Garc{\'\i}a-Ojalvo
 for an  interesting 
discussion on the problem studied here, as well as
to 
 Raul Toral, Pedro Garrido,
  and Luciano Pietronero
 for useful comments.
This work has been  
supported by the European Union through a grant to M.A.M.
(contract ERBFMBICT960925) and by Direcci\'on General de Investigaci\'on
Cient\'ifica y T\'ecnica, Project, PB96-0241.

\end{text}


\begin{references}
\bibitem{note}
 We have not succeed so far in relating $\mu$
to other known exponents using scaling arguments.
\bibitem{first} We cannot rigorously exclude the possibility of 
the NIDT being discontinuous (first order).
\end{references}
\end{document}